\documentclass[aip,jap,a4,reprint,twoside,floatfix,superscriptaddress]{revtex4-1}
\usepackage{amssymb}

\usepackage{graphicx}
\usepackage{dcolumn}
\usepackage{bm}
\usepackage{epsfig}
\newcommand{\ignore}[1]{}

\makeatletter
\newcommand{\thickhline}{%
    \noalign {\ifnum 0=`}\fi \hrule height 1.2pt
    \futurelet \reserved@a \@xhline
}
\newcolumntype{"}{@{\hskip\tabcolsep\vrule width 2pt\hskip\tabcolsep}} \makeatother

\begin{document}


\title{Effects of Li doping on H-diffusion in MgH$\mbox{\boldmath $_2$}$: a first-principles study}

\author{Wenmei Ming}
\affiliation{Department of Materials Science and Engineering
University of Utah, Salt Lake City, UT 84112, USA}

\author{Zhigang Zak Fang}
\affiliation{Department of Metallurgical Engineering, University of
Utah, 135 South 1460 East, Room 412, Salt Lake City, Utah 84112-0114
}

\author{Feng Liu}
\thanks{Corresponding author. E-mail: fliu@eng.utah.edu}
\affiliation{Department of Materials Science and Engineering
University of Utah, Salt Lake City, UT 84112, USA}




\begin{abstract}
The effects of Li doping in MgH$_2$ on H-diffusion process are
investigated, using first-principles calculations. We have
identified two key effects: (1) The concentration of H vacancy in
the $+1$ charge state (V$_H^{+1}$) can increase by several orders of
magnitude upon Li doping, which significantly increases the vacancy
mediated H diffusion rate. It is caused by the preferred charge
states of substitutional Li in the $-1$ state (Li$_{Mg}^{-1}$) and
of interstitial Li in the $+1$ state (Li$_i^{+1}$), which indirectly
reduce the formation energy of V$_H^{+1}$ by up to 0.39 eV depending
on the position of Fermi energy. (2) The interaction between
V$_H^{+1}$ and Li$_{Mg}^{-1}$ is found to be attractive with a
binding energy of 0.55 eV, which immobilizes the V$_H^{+1}$ next to
Li$_{Mg}^{-1}$ at high Li doping concentration. As a result, the
competition between these two effects leads to large enhancement of
H diffusion at low Li doping concentration due to the increased
H-vacancy concentration, but only limited enhancement at high Li
concentration due to the immobilization of H vacancies by too many
Li.
\end{abstract}

\maketitle

\section{Introduction}
Light metal hydride MgH$_2$ is one of the most promising
hydrogen-storage materials for on-board clean-fuel application,
because it has both high gravimetric (7.7wt\%) and volumetric
densities (6.7$\times$ 10$^{22}H/cm^3$) \cite{Hdensity}. However,
their dehydrogenation process is too slow to be practically useful,
and for bulk materials as high as about 300 K above room temperature
is required to obtain an equilibrium H$_2$ pressure of 1 bar
\cite{Hdensity, PressureTemp1, PressureTemp2, PressureTemp3}. Such
poor dehydrogenation kinetics is primarily due to the strong ionic
bonding between Mg and H and large enthalpy of formation of MgH$_2$
($\thicksim$ 75KJ/molH$_2$), as evidenced by both experiments
\cite{bondE1, bondE2, bondE3} and first-principles calculations
\cite{bondT1, bondT2}. Various attempts have been made to help
facilitate dehydrogenation process. For example, to improve the
kinetics, ball milling processing \cite{ball1, ball2} has been used
to shorten the diffusion length, doping of transition metals
\cite{ball1, ball2, dop1} were adopted to reduce the strength of
H-Mg bond, and applying tensile stress was tried to weaken the Mg-H
stability \cite{PressureTemp1, stress1}.

On the other hand, doping has been known to enhance H
diffusion in metal hydrides, which is usually mediated by H vacancy, by inducing a higher concentration of H-vacancy
\cite{dopinduce1, dopinduce2, dopinduce3, dopinduce4, dopinduce5,
dopinduce6}. For example, Van de Walle \emph{et. al} recognized that in certain charged
state, Zr(Ti) can enhance the dehydrogenation kinetics of NaAlH$_4$
\cite{dopinduce1}, because the formation energy of
H-vacancy is decreased upon doping. In particular, when the H-vacancy is charged, its formation energy
depends on the position of Fermi energy; and conversely, selective doping of
the hydride with impurities that take different charged states will tune the Fermi
energy with respect to the dopant-free system. And the shift of
Fermi energy can result in a decrease of H-vacancy formation
energy depending on the sign of the H-vacancy charge state. Consequently, the
concentration of H-vacancy will increase to enhance the vacancy-mediated H diffusion.

In this work, we investigated the effects of Li doping on H
diffusion in MgH$_2$. One important reason that we chose Li is
because it is a lighter metal than Mg, so that it will not degrade
the high H gravimetric density. We focused on the effects of charge
state of Li impurity and H-vacancy, as recognized before in other
systems\cite{dopinduce1}, but also went beyond the previous works by
taking into account the effects of interaction between the charged
impurities and defects. In many previous studies of the
charged-impurity-enhanced H diffusion\cite{dopinduce1, dopinduce2,
dopinduce3, dopinduce4, dopinduce5, dopinduce6}, it implicitly
assumed no interaction between the dopant and defect. This might be
true in the limit of low doping concentration and weak defect-dopant
interaction, but unlikely at high doping concentration. Especially,
if there is an attractive impurity-defect interaction, such as the
binding between the Li-dopant and H-vacancy in MgH$_2$ as shown by
Smith \emph{et al.} \cite{Livac}, the impurity may immobilize the
H-vacancy, counteracting the enhancement effect of H-vacancy on H
diffusion.

Therefore, by taking into account the binding between Li and
H-vacancy and its dependence on the charge states of Li and
H-vacancy, we have systematically studied the effects of Li doping
on H diffusion in MgH$_2$ as a function of Li concentration. We have
determined the favored charge states of Li by calculating its
formation energy as a function of Fermi energy, the equilibrium
concentration of H vacancies by calculating the H vacancy formation
energy as a function of Li doping concentration, and the percentage
of immobilized H-vacancies by calculating the binding energies
between H-vacancy and Li-dopant. We have also calculated the
diffusion barrier of H-vacancy in the presence of Li dopant.

\section{Calculation details}
Our first principles calculations based on density functional theory
(DFT) were conducted using projector augment wave pseudopotential
(PAW) \cite{PAW} with the generalized gradient approximation (GGA)
\cite{GGA} to the exchange-correlation functional, as implemented in
VASP package \cite{vasp}. Supercell technique was used to calculate
the formation energy of defects and dopants, interaction energy and
diffusion barrier. We used a supercell comprised of
$3\times3\times4$ primitive MgH$_2$ rutile unit cells with the
dimensions of $13.481\times13.481\times12.012 {\AA}^3$. 400 eV
energy cutoff and $2\times2\times2$ k-mesh were used for wavefuntion
expansion and k-space integration, respectively. All the structures
were relaxed in terms of internal atomic coordinates using conjugate
gradient method until the force exerted on each atom was smaller
than $0.005eV/{\AA}^3$. The charged system was simulated by adding
to or removing from the system electrons with a compensating uniform
opposite charge background. Diffusion barrier was calculated using
the nudged elastic band method \cite{NEB}.

The formation energy $E^{f}(X^q)$ of defect or dopant ($X$) with
charge q was computed according to Ref.\cite{chargecal}:
\begin{eqnarray}
\Delta E^{f}(X^q)=E_{tot}(X^q)-E_{tot}(bulk)-\sum_i n_i\mu_i
\nonumber \\
 +q(E_F+E_{\nu}+\Delta V).
\end{eqnarray}
Where $E_{tot}(bulk)$ and $E_{tot}(X^q)$ are the total energies of
supercell for pure MgH$_2$ and for MgH$_2$ containing defect or
dopant ($X^q$), respectively. $E_{\nu}$ is chosen to be the valence
band maximum (VBM) energy. $E_{F}$ is the Fermi energy with respect to
$E_{\nu}$. $\Delta V$ is additional electrostatic energy alignment
due to different energy references between the defect-containing structure
and defect-free structure. $i$ denotes H-defect or dopant Li and $n_i$ is
the number of species $i$ in the supercell.
$\mu_i$ is the chemical potential of species $i$. In low
concentration limit, the equilibrium defect concentration can be related to the formation energy using:
\begin{eqnarray}   \label{concentration}
C=N\exp(-\Delta E^f/k_BT)
\end{eqnarray}
N is the number of sites that can be occupied by defect, $k_B$ is
Boltzman constant and T is temperature in K.

For the chemical potential $\mu_i$, the externally added dopant Li is
assumed to have its bulk chemical potential $E_{Li}(bulk)$. The chemical potential
of H, $\mu_H$ is in between
$\frac{1}{2}E(H_2)+\frac{1}{2}\Delta H_f(MgH_2)$ (H-poor condition)
and $\frac{1}{2}E(H_2)$ (H-rich condition), considering
thermodynamic equilibrium between MgH$_2$, Mg and H$_2$. $\Delta
H_f(MgH_2)$ is enthalpy of formation of MgH$_2$, E(H$_2$) is the
energy of hydrogen molecule at 0 K. Similarily, the chemical
potential of Mg is in between E(bulk Mg) and E(bulk
Mg)+$\Delta$H$_f$(MgH$_2$). We specifically considered two extreme
cases: H poor condition and H rich condition.

\section{Results and discussion}
First we calculated the formation energy of native defects:
H-vacancy (V$_H$ with charge -1, 0, +1) and interstitial H (H$_i$
with charge -1, 0, +1). The preferred defects are $V_H^{+1}$ and
$V_H^{-1}$ in H-poor condition, and $V_H^{+1}$ and $H_i^{-1}$ in
H-rich condition, respectively. Charge neutrality condition requires
Fermi-energy to be 2.85 eV and 2.65 eV for the H-poor condition and
H-poor condition, respectively. These results are in good agreement
with those for MgH$_2$ in Ref. \cite{dopinduce4}. In Table
\ref{tablec} we give an estimate of the concentration for the
favored H-defects from equation (\ref{concentration}).
\begin{figure}
\includegraphics[clip,scale=0.45]{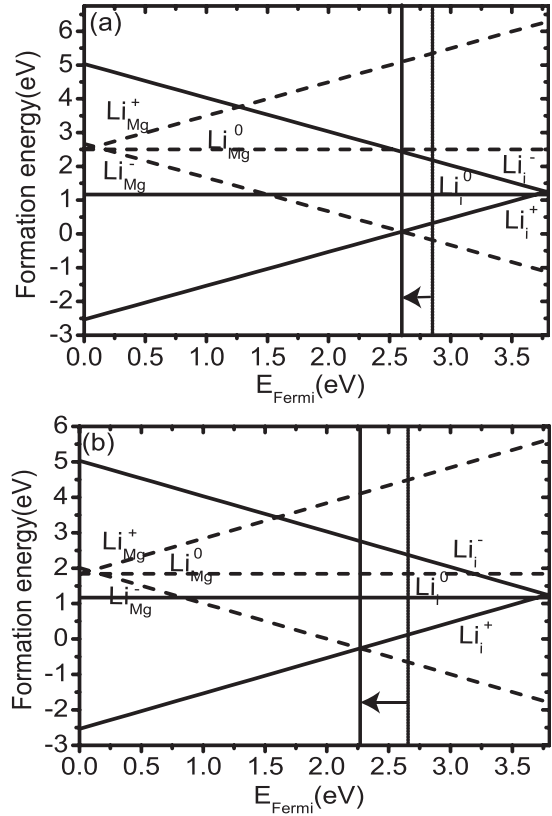}
\caption{\label{Fig1}Formation energy of Li-dopant in MgH$_2$:(a)
H-poor condition. (b) H-rich condition. The vertical solid and
dashed lines indicate the Fermi energy in MgH$_2$ with and
without Li, respectively. E$_{Fermi}$=0 eV corresponds to the VBM and E$_{Fermi}$=3.8 eV corresponds to the CBM.}
\end{figure}

In order to study how the formation energies of the H-related
defects are affected by Li doping, we then calculated the formation
energy for both substitutional Li configuration (Li$_{Mg}$) and
interstitial Li configuration (Li$_i$) in the (-1, 0, +1) charge
states. As shown in Figure \ref{Fig1} for both H-poor and H-rich
conditions, Li$_{Mg}^{-1}$ is more stable than Li$_{Mg}^{0}$ and
Li$_{Mg}^{+1}$ in almost the whole range of Fermi energy in the gap
except very close to the VBM. While Li$_i^{+1}$ is more stable than
other two charge states in almost the whole range of Fermi energy in
the gap except very close to conduction band minimum (CBM). This
indicates that the defect level remains close to the VBM and CBM for
Li$_{Mg}$ and Li$_i$, respectively (see Ref. \cite{TiO2} for similar
behavior of native defects in anatase TiO$_2$). Under the
charge-neutrality condition, the Fermi energy of the Li-doped system
(vertical solid lines) is shifted to the left by 0.25 eV (Fig. 1(a))
and 0.39 eV (Fig. 1(b)) with respect to the Fermi energy of the
undoped system (vertical dashed lines) for the H-poor and H-rich
condition, respectively, assuming the concentration of dopant Li is
much higher than that of H-defects so that the Li$_i^{+1}$ and
Li$_{Mg}^{-1}$ are the dominant charged dopants to maintain the
charge-neutrality condiction. The Fermi energy in both situations is
deep inside the band gap. Thus, the thermally excited free carriers
in both valence and conduction band are negligible. The consequence
of the shift of Fermi energy is that the formation energy of
V$_H^{+1}$ is reduced by 0.25 eV and 0.39 eV under the H-poor and
H-rich condition, respectively, according to Eq. (1). And the
opposite effect happens to V$_H^{-1}$ and H$_i^{-1}$: their
formation energy is increased by 0.25 eV and 0.39 eV, respectively.
As shown in Table \ref{tablec}, at 400K the concentration of
V$_H^{+1}$ in the Li-doped system is 1.40$\times$10$^3$ and
8.12$\times$10$^4$ times larger than that in the undoped system
under the H-poor and H-rich condition, respectively. On the
contrary, the concentration of H$_i^{-1}$ in the Li-doped system is
$\thicksim$10$^{5}$ and $\thicksim5\times10^{7}$ times lower than
that in the undoped system under the H-poor and H-rich condition,
respectively.

A previous calculation\cite{dopinduce5} showed that in the undoped
MgH$_2$ the diffusion barrier of $V_H^{+1}$ is 0.25 eV smaller than
that of $V_H^{-1}$ under the H-poor condition, and the diffusion
barrier of $V_H^{+1}$ is 0.36 eV higher than that of H$_i^{-1}$
under the H-rich condition. This means that without Li doping, the
$V_H^{+1}$ is the dominant diffusing species under the H-poor
condition, while the H$_i^{-1}$ is the dominant diffusing species
under the H-rich condition. Our calculations show that upon Li
doping, the formation energy of $V_H^{+1}$ is decreased by 0.25 eV
under that H-poor condition, while that of H$_i^{-1}$ is increased
by 0.39 eV under H-rich condition. Because the H-related defect
diffusion is determined by the activation barrier, which is the sum
of the diffusion barrier and the formation energy. The $V_H^{+1}$
remains the dominant diffusing species under the H-poor condition
because its formation energy is decreased, leading to a lower
activation barrier. In contrast, the H$_i^{-1}$ becomes the less
favorable diffusing species under the H-rich condition because its
formation energy is increased, leading to a higher activation
barrier. Consequently, the Li doping makes the $V_H^{+1}$ the
dominant diffusion species in the whole range of H chemical
potential.

We note that we have neglected entropy contribution in our analysis.
Usually, this is a good approximation because the contribution due
to the entropy difference is much smaller than the contribution due
to the total energy difference. Of course, more accurate results can
be obtained by calculating the phonon spectra of all the MgH$_2$
systems and H$_2$. On the other hand, for the MgH$_2$ system we
consider, it has been shown that even though H has a low mass, but
the vibrational entropies for H in the lattice and in the H$_2$
reservoir are rather similar and hence the net entropy difference is
small \cite{dopinduce1}. Also we have used a relatively large
supercell dimension so that the added defect charge density in the
supercell is very low. Consequently, the interaction energy between
the charged defects in the neighboring cells is expected to be
sufficiently small, not to affect our conclusion.

\begin{table}
\caption{\label{tablec} The formation energy($\Delta$ E$^f$) and
concentration(C) of relevant H-defects without(a) and with(b) dopant
Li, at T$=$400 K.}
\begin{tabular} {l|c|c|c|c}
\hline
& \multicolumn{2}{c|}{H-poor} &\multicolumn{2}{c}{H-rich}\\
\thickhline
&V$_H^{+}$ &V$_H^{-}$ &V$_H^{+}$ &H$_i^{-}$ \\
\hline
$\Delta$ E$^f$(eV)$^a$ &1.225 &1.225 &1.358 &1.358     \\
\hline
C($/cm^3$)$^a$&2.5$\times$10$^7$& 2.5$\times$10$^7$ & 5.3$\times$10$^5$ & 5.3$\times$10$^5$   \\
\thickhline
$\Delta$ E$^f$(eV)$^b$ &0.975 &1.475 &0.968 &1.748    \\
\hline
C($/cm^3$)$^b$& 3.5$\times$10$^{10}$& 1.772$\times$10$^{4}$ &4.13$\times$10$^{10}$ &6.5    \\
\cline{1-5}
\end{tabular}
\end{table}

\begin{figure}
\includegraphics[clip,scale=0.65]{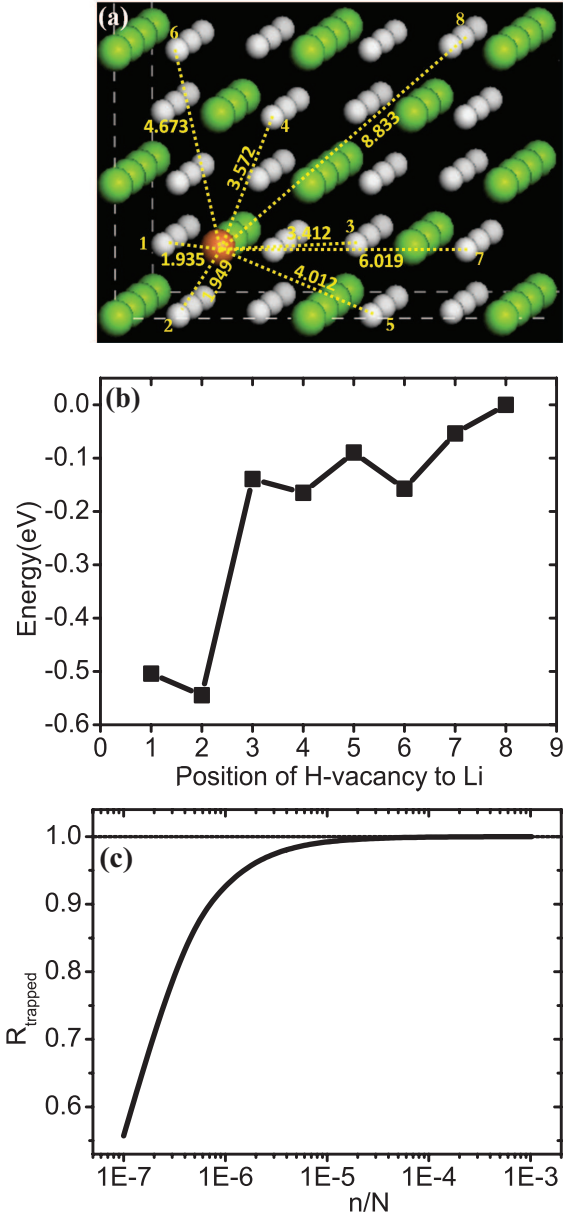}
 \caption{\label{binding} (color online) (a) The structure of Li-dopant plus H-vacancy with H-vacancy at different positions labeled with number and distance from Li; (b)Interaction energy between V$_H^{+}$ and Li$_{Mg}^{-1}$ as a function of their separation
 distance (in Angstrom); (c) Ratio of the trapped V$_H^{+}$ with Li$_{Mg}^{-1}$
 to the number of V$_H^{+}$, T=400 K. Green balls are Mg atoms, white balls are
H atoms and orange ball is Li dopant.}
\end{figure}

The results above suggest the dominating defect and dopant species
to be V$_H^{+}$, Li$_{Mg}^{-1}$ and Li$_i^{+1}$. However, we didn't
consider the interaction between V$_H^{+}$ and Li$_{Mg}^{-1}$. Next,
we calculated the attractive interaction energy between V$_H^{+}$
and Li$_{Mg}^{-1}$ as a function of their separation as shown in
Fig. \ref{binding}. We didn't consider the interaction between
V$_H^{+}$ and Li$_i^{+1}$ because it is repulsive. Two key features
are found in Fig. \ref{binding}(b): (1) V$_H^{+}$ prefers to sit in
one of the six nearest-neighbor H-sites (site 1 and site 2 in Fig.
\ref{binding}(a)) of Li with binding energy of 0.50-0.55 eV; (2)
Once beyond the nearest-neighbor H-site, their attraction decays
rapidly to be insignificant. Based on this observation, we propose a
nearest-neighbor interaction model to determine how many V$_H^{+}$
being trapped by Li$_{Mg}^{-1}$ as a function of Li doping
concentration. We assume that the interaction energy is
$\Delta$E$_b$ = -0.55 eV when V$_H^{+}$ is in any of the six
nearest-neighbor sites and negligible otherwise. Following the
Boltzmann distribution \cite{BoZ} we have
\begin{eqnarray}
R_{trapped}=\frac{3n\exp[-\Delta E_b/k_BT]}{[2N-3n]+3n\exp[-\Delta
E_b/k_BT]}.
\end{eqnarray}
Where R$_{trapped}$ is the ratio of the number of trapped V$_H^{+}$
to the total number of V$_H^{+}$, n is the number of doped Li and N
is the number of Mg sites. The number of substitutional and interstitial Li are taken to the same under
the charge-neutrality condition, as shown in Fig. \ref{Fig1}.

Fig. \ref{binding}(c) shows the calculated R$_{trapped}$ as a
function of Li doping concentration. We see that even in the low
concentration (for example, $\frac{n}{N}=1\times10^{-4}$), the
trapping ratio R$_{trapped}$ is very close to one, indicating that
almost all the V$_H^{+}$ next to Li are immobilized due to their
attractive interaction. This also indicates that H vacancy prefers
to stay next to Li$_{Mg}^{-1}$, because its formation energy is
effectively decreased by 0.55 eV.

\begin{figure}
\includegraphics[clip,scale=0.50]{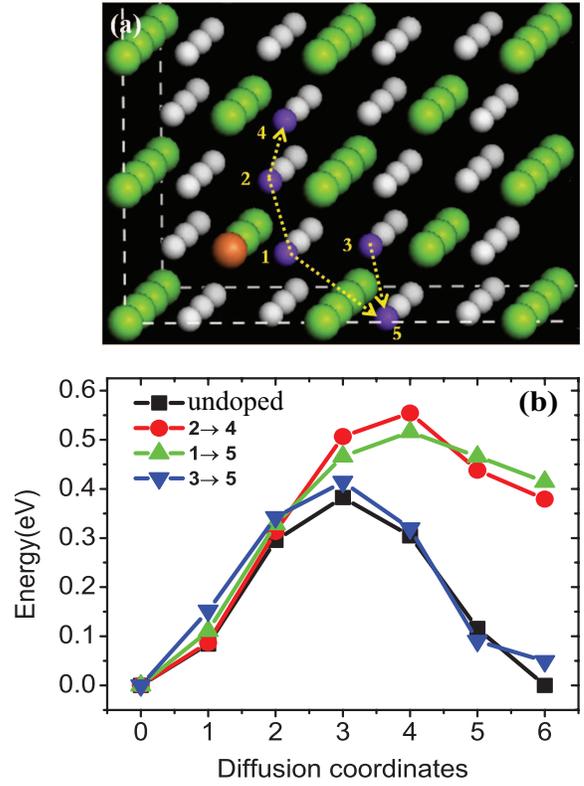}
\caption{(color online) (a) Illustration of different diffusion path
for V$_H$ to diffuse away further way from Li site. The arrow
indicates diffusion direction; (b) V$_H$ mediated H-diffusion
barrier change with the presence of Li. The purple ball indicates
the position of V$_H$ in our calculation. The arrow indicates the
diffusion path}
\end{figure}

Furthermore, we studied kinetically how H-vacancy diffusion is
affected by the presence of Li$_{Mg}$ through the calculation of diffusion
barriers. In Fig. 3, we show the barriers for the H-vacancy diffusing from the nearest-neighbor sites of Li (sites 1 and 2 in Fig. 3(a) ) to its closest
H site (sites 4 and 5) (path 1) and from the next nearest-neighbor site
(site 3) to its closet H site (site 5) (path 2). For the path 1, the diffusion barrier is found to increase by 0.15 eV
compared to that in the undoped MgH$_2$. For the path 2, the diffusion barrier is found only
$\sim$30 meV higher than that in the undoped MgH$_2$. This strong site
dependence of H-vacancy diffusion barrier is consistent with the fast
decay of the attractive interaction between V$_H$ and
Li$_{Mg}$ as shown in Fig. 2(b). The 0.15 eV increase of diffusion barrier,
together with the large V$_H$ trapping ratio suggest that H vacancies will mostly
be immobilized in the vicinity of Li dopants, inhibiting the V$_H$ mediated H diffusion.

\section{Conclusions}
In conclusion, we have investigated the effects of Li-doping in
MgH$_2$ on the H-vacancy meditated H-diffusion, using DFT
calculations. The formation energy calculation shows that the Li
dopant favors two charged configurations of Li$_{Mg}^{-1}$ and
Li$_i^{+1}$. The charge neutrality condition requires the Fermi
energy be shifted towards the VBM by 0.25 eV and 0.39 eV upon Li
doping under the H-poor and H-rich conditions, respectively, which
decreases the formation energy of V$_H^{+}$ by the same amount. This
leads to an increase of V$_H^{+}$ concentration by up to about 5
orders of magnitude at T=400 K. Furthermore, the calculations of
interaction energy between V$_H^{+1}$ and Li$_{Mg}^{-1}$ as well as
diffusion barrier of H vacancy in the presence of Li show that
almost all the H-vacancy next to Li are immobilized. Therefore, the
H-diffusion is enhanced by Li doping in MgH$_2$ only at the low Li
doping concentration but not at the high concentration.

\section{Acknowledgement}
This work was supported by NSF MRSEC (Grant No. DMR-1121252) and
DOE-BES (Grant No. DE-FG02-04ER46148). We thank the CHPC at
University of Utah and NERSC for providing the computing resources.

\end{document}